\documentclass[journal]{IEEEtran}
\usepackage{amsmath,amsfonts}
\usepackage{amsthm}
\theoremstyle{definition}
\newtheorem{definition}{Definition}[section]
\theoremstyle{plain}
\newtheorem{proposition}{Proposition}[section]
\theoremstyle{definition}
\newtheorem{example}{Example}[section]
\usepackage{algorithmic}
\usepackage{algorithm}
\usepackage{array}
\usepackage[caption=false,font=normalsize,labelfont=sf,textfont=sf]{subfig}
\usepackage{textcomp}
\usepackage{stfloats}
\usepackage{url}
\usepackage{verbatim}
\usepackage{graphicx}
\usepackage{cite}
\def\BibTeX{{\rm B\kern-.05em{\sc i\kern-.025em b}\kern-.08em
    T\kern-.1667em\lower.7ex\hbox{E}\kern-.125emX}}
\usepackage{balance}

\usepackage{multirow}
\usepackage{booktabs}
\usepackage{siunitx}
\usepackage{longtable}

\usepackage[hidelinks]{hyperref}

\newcommand{\nop}[1]{}

\usepackage{xcolor}
\newcommand{\revise}[1]{#1}
\newenvironment{revisionblock}
{\begingroup}
{\endgroup}

\begin{document}
\title{Shapley Value on Uncertain Data}
\author{Zhuofan Jia and Jian Pei\\
Duke University, Durham, NC, USA\\
\texttt{\{zhuofan.jia, j.pei\}@duke.edu}}

\maketitle

\begin{abstract}
The Shapley value provides a principled framework for allocating rewards according to participants' contributions. Existing data-valuation methods typically assume that each participant contributes a fixed dataset. In practice, however, data are often provided as random samples from underlying distributions, making the resulting Shapley value a random variable. We formulate the Shapley value for probabilistic data distributions and characterize its expectation and variance to quantify both the average contribution and its sampling uncertainty. We derive closed-form expressions whenever possible and unbiased empirical estimators, together with statistical analyses of their properties. We further develop three Monte Carlo estimation methods: a baseline method based on independent sampling, a pooled method that reduces source-data access through sample reuse, and a stratified pooled method that adaptively allocates the sampling budget according to player-specific variability. Experiments on synthetic and real datasets demonstrate that the proposed methods accurately estimate both moments of the probabilistic Shapley value, while the pooled methods substantially improve estimation stability under constrained source-data access. This work extends Shapley value analysis from deterministic datasets to probabilistic data distributions, providing a practical framework for uncertainty-aware data valuation.
\end{abstract}


\section{Introduction}

The Shapley value~\cite{shapley1953value}, introduced in cooperative game theory, provides an axiomatic framework for allocating value according to participants' contributions. Its fairness guarantees---efficiency, symmetry, null-player, and additivity~\cite{shapley1954method,winter2002shapley,peleg2007cooperative,chalkiadakis2011computational}---have led to widespread applications in economics~\cite{shapley1953value}, political science~\cite{shapley1954method}, and, more recently, machine learning~\cite{strumbelj2010efficient,lundberg2017unified,aas2021explaining,kumar2021limits,ghorbani2019data}. In collaborative learning, it provides a principled mechanism for quantifying each data provider's contribution to model performance~\cite{ghorbani2019data,jia2019towards}.

Most existing data-Shapley methods assume that each participant contributes a fixed dataset~\cite{ghorbani2019data,jia2019towards}. Under this assumption, the Shapley value is computed or approximated by evaluating model performance on coalitions of static datasets~\cite{deng1994complexity,castro2009random,maleki2013bounding,mitchell2022sampling,burgess2021bernstein,illes2019ergodic}. In many applications, however, participants provide only \emph{samples} drawn from underlying, often unknown, data distributions, such as in data markets, privacy-preserving systems, and federated learning~\cite{ghorbani2020distributional,kwon2021aistats,cai2024chgshapley}. \revise{Each realization of player samples defines a deterministic cooperative game with deterministic marginal contributions. Across repeated realizations, however, the sampled datasets and resulting marginal contributions vary. We refer to this variability as \emph{sampling-induced uncertainty}.}

Sampling-based data sharing is often necessary in practice. In cross-silo federated learning across hospitals or financial institutions, regulations such as HIPAA and GDPR limit the amount of patient or customer data transmitted in each communication round~\cite{mcmahan2017fedavg,kairouz2021flsurvey,li2020flchallenges,bao2023fedcollab}. Institutions therefore share only randomly selected subsets of local data to satisfy privacy constraints and reduce communication costs~\cite{li2025healthcareflreview,madathil2025healthcareflsystems,dwork2014dpbook,abadi2016dpsgd}. \revise{Consequently, different training rounds may use different data realizations. Although the marginal contribution is deterministic for a realized dataset, it varies across rounds because the contributed data are randomly sampled}~\cite{liu2022neurrips_crosssilo_dp,liu2024rpdpfl}.

\revise{Consider a data market in which hospitals provide access to de-identified patient records for developing a hospital-readmission model. Rather than transferring complete databases, each hospital provides a randomly sampled batch of 500 records per query because of privacy, licensing, and access-budget constraints~\cite{pei2020survey,kennedy2022datamarkets,zhang2023survey,hao2023evolution}. The market operator evaluates model performance for different provider coalitions and allocates compensation using the Shapley value. Because repeated evaluations may use different sampled batches, coalition utilities and Shapley values vary across data realizations, although each realized valuation is deterministic. The expected Shapley value measures a provider's average contribution, while its variance quantifies the stability of that contribution across repeated data purchases.}

\revise{These examples show that sampling-based data access is a practical consequence of privacy, licensing, communication, and access-budget constraints. Reliable valuation under such settings therefore requires a probabilistic formulation that characterizes both the expectation and variance of each provider's Shapley value~\cite{ghorbani2020distributional,kwon2021aistats}.}

\revise{Sampling-induced uncertainty makes the Shapley value a random variable across data realizations, although it remains deterministic for any realized dataset. For example, two medical institutions may each hold one million patient records but share only 500 samples per round. Repeating the valuation with newly sampled records produces different Shapley values, whose distribution motivates estimating both their expectation and variance~\cite{owen2013monte,asmussen2007stochsim,glasserman2004monte}.}

\revise{Estimating this variance benefits data owners, buyers, and market operators. Owners can assess the stability of their valuations, buyers can distinguish persistent contribution from sampling fluctuations, and operators can design more transparent pricing mechanisms~\cite{zhang2023survey,kennedy2022datamarkets}. Existing data-valuation methods do not jointly estimate the expectation and variance of provider-level Shapley values induced by repeated random data access.}

This work fills this gap by introducing a framework for the Shapley value of \emph{probabilistic data distributions}. We treat the Shapley value as a random variable induced by sampling from players' distributions and characterize its expectation and variance whenever closed-form solutions exist. This formulation captures both the average contribution of each player and the uncertainty introduced by stochastic sampling, extending deterministic Shapley valuation to probabilistic data.

The main contributions are as follows.

\begin{itemize}
    \item \textbf{A probabilistic formulation of data Shapley.}
    We formulate the Shapley value when each participant contributes samples from an underlying distribution, making the Shapley value a random variable. Its expectation and variance characterize the average contribution and its stability.

    \item \textbf{Closed-form characterization of the Shapley expectation and variance.}
    We derive analytical expressions for the expectation and variance whenever the underlying marginal contributions admit closed-form expressions, establishing a theoretical foundation for uncertainty-aware data valuation.

    \item \textbf{Unbiased empirical estimators.}
    \revise{We derive unbiased outer-sample estimators for the expectation and variance when the Shapley value of each realized game is computed exactly. We further separate sampling-induced uncertainty from the additional error introduced by finite permutation sampling.}

    \item \textbf{Three scalable Monte Carlo estimation algorithms.}
    We develop (i) a \emph{baseline} estimator based on independent sampling,
    \revise{(ii) a \emph{pooled} estimator that reduces fresh-source access through sample reuse, and (iii) a \emph{stratified pooled} estimator that reallocates a fixed local sampling budget according to player-specific variability.}

    \item \textbf{Comprehensive empirical evaluation.}
    \revise{We evaluate regression and nonlinear classification tasks, sensitivity to the pooling parameters, scalability to \(1{,}000\) players, variance efficiency under matched source budgets, and runtime under simulated source latency. The results show that pooling is most effective when source access is constrained, trading additional local computation for fewer source requests and more stable estimates.}
\end{itemize}

Together, these contributions provide a unified framework for uncertainty-aware data valuation by quantifying both the expectation and variance of probabilistic Shapley values.

The remainder of this paper is organized as follows. Section~\ref{sec:problem} formulates the problem and defines the Shapley value for probabilistic data distributions. Section~\ref{sec:theory} presents the theoretical analysis. Section~\ref{sec:empirical} develops the empirical estimation methods and reports the experimental results. Section~\ref{sec:related} reviews related work, and Section~\ref{sec:conclusion} concludes the paper.

\section{Problem Formulation}\label{sec:problem}

\revise{We formalize Shapley value estimation when participants contribute samples from underlying data distributions. Each realization defines a deterministic cooperative game, and randomness arises only across realizations, where different sampled datasets induce different coalition utilities, marginal contributions, and Shapley values. We first review the classical Shapley value and then define its expectation and variance under this sampling process.}

\subsection{Shapley Value and Computation}\label{sec:shapley}

Consider a cooperative game with player set \(N=\{1,2,\ldots,n\}\) and utility function \(v:2^N\rightarrow\mathbb{R}\), where \(v(S)\) denotes the value of coalition \(S\subseteq N\). The objective is to allocate \(v(N)\) among the players according to their marginal contributions.

\begin{definition}[Shapley value \cite{shapley1953value}]
The Shapley value of player \(i\in N\) is
\begin{equation}
\phi_i=\sum_{S\subseteq N\setminus\{i\}}
\frac{|S|!(n-|S|-1)!}{n!}
\bigl(v(S\cup\{i\})-v(S)\bigr),
\label{eq:shapley_deterministic}
\end{equation}
where \(v(S\cup\{i\})-v(S)\) is the marginal contribution of player \(i\) to coalition \(S\).
\end{definition}

The Shapley value satisfies four fundamental properties~\cite{shapley1953value,shapley1954method,winter2002shapley,peleg2007cooperative,chalkiadakis2011computational}: \emph{efficiency}, \emph{symmetry}, the \emph{null-player} property, and \emph{additivity}. Specifically,
\(\sum_{i\in N}\phi_i(v)=v(N)\); players with identical marginal contributions receive equal values; players with zero marginal contribution receive zero value; and \(\phi_i(v+w)=\phi_i(v)+\phi_i(w)\).

An equivalent formulation averages marginal contributions over all player permutations~\cite{shapley1953value}. Let \(\Pi(N)\) denote the set of permutations of \(N\), and let \(P_i^\pi\) be the players preceding \(i\) in permutation \(\pi\). Then
\begin{equation}
\phi_i=\frac{1}{n!}\sum_{\pi\in\Pi(N)}
\bigl(v(P_i^\pi\cup\{i\})-v(P_i^\pi)\bigr).
\label{eq:perm}
\end{equation}

Equation~\eqref{eq:perm} expresses the Shapley value as the average marginal contribution over all player orderings. Enumerating all permutations is computationally infeasible for large \(n\): exact computation is \#P-hard for general cooperative games~\cite{deng1994complexity} and remains intractable for compact representations such as MC-nets~\cite{elkind2009computational}. A standard Monte Carlo approximation samples \(k\) random permutations \(\{\pi_1,\ldots,\pi_k\}\subseteq\Pi(N)\):
\begin{equation}
\widehat{\phi_i(v)}
=\frac{1}{k}\sum_{j=1}^{k}
\bigl(v(P_i^{\pi_j}\cup\{i\})-v(P_i^{\pi_j})\bigr).
\label{eq:mc}
\end{equation}

\revise{Equation~\eqref{eq:mc} follows the classical permutation-sampling framework of Castro et al.~\cite{castro2009random}, which approximates the Shapley value of a fixed cooperative game by sampling player orderings. In our framework, it serves as the inner estimator for a realized game, conditioned on the sampled player datasets.}

In collaborative machine learning, each player contributes a dataset, and \(v(S)\) measures the performance of a model trained on the data contributed by coalition \(S\). The Shapley value therefore quantifies each dataset's contribution to model performance.

\subsection{Shapley Value of Probabilistic Distributions}\label{sec:probshapley}

\revise{We extend the classical formulation by letting player \(i\) own a data distribution \(P_i\) rather than a fixed dataset. A realization of \(\{P_i\}_{i=1}^{n}\) defines a deterministic cooperative game and Shapley value. Repeated sampling induces a distribution over realized games and thus a random Shapley value for each player.}

Our goal is to estimate the expectation \(\mathbb{E}[\phi_i(v)]\) and variance \(\mathrm{Var}(\phi_i(v))\) of each player's Shapley value. They quantify the player's average contribution and its sampling uncertainty, respectively, enabling uncertainty-aware data valuation. 

\subsection{The Baseline Method}\label{sec:baseline}

\begin{algorithm}[!t]
\caption{\textsc{MC-Shapley}: Monte Carlo Shapley Value Estimation for a Realized Game}
\label{alg:mc_shapley}
\begin{algorithmic}[1]
\REQUIRE Realized player datasets $\{D_i^{(g)}\}_{i=1}^n$, number of iterations $n_{\text{iter}}$
\ENSURE Approximate Shapley values $\{\bar{\phi}_i^{(g)}\}_{i=1}^n$
\STATE Initialize $\bar{\phi}_i^{(g)} \gets 0$ for all $i$
\FOR{$t \gets 1$ to $n_{\text{iter}}$}
    \STATE Sample permutation $\pi \sim \text{Uniform}(\Pi(N))$
    \STATE $S \gets \emptyset$, $v_{\text{prev}} \gets 0$
    \FORALL{$i$ in order $\pi$}
        \STATE $D_{S \cup \{i\}}^{(g)} \gets \bigcup_{j \in S \cup \{i\}} D_j^{(g)}$
        \STATE $v_{\text{curr}} \gets v(S \cup \{i\} \mid D_{S \cup \{i\}}^{(g)})$
        \STATE $\Delta_i \gets v_{\text{curr}}-v_{\text{prev}}$
        \STATE $\bar{\phi}_i^{(g)} \gets \bar{\phi}_i^{(g)}+\Delta_i$
        \STATE $S \gets S \cup \{i\}$, $v_{\text{prev}} \gets v_{\text{curr}}$
    \ENDFOR
\ENDFOR
\FOR{$i \gets 1$ to $n$}
    \STATE $\bar{\phi}_i^{(g)} \gets \bar{\phi}_i^{(g)}/n_{\text{iter}}$
\ENDFOR
\RETURN $\{\bar{\phi}_i^{(g)}\}_{i=1}^n$
\end{algorithmic}
\end{algorithm}

\revise{We propose a Monte Carlo framework for estimating Shapley values under uncertain data. For each realized game, the \textsc{MC-Shapley} subroutine (Algorithm~\ref{alg:mc_shapley}) estimates the per-game Shapley values \(\bar{\phi}_i^{(g)}\) by averaging marginal contributions over \(n_{\text{iter}}\) random permutations. The baseline method (Algorithm~\ref{alg:baseline}) applies this subroutine to independently sampled games \(\{D_i^{(g)}\}_{i=1}^n\).}

For each of the \(n_{\text{games}}\) games, we independently sample
\begin{equation}
D_i^{(g)}
=
\{(x_j,y_j)\sim P_i\}_{j=1}^{n_{\text{sample}}},
\qquad
i=1,\ldots,n.
\label{eq:sampling}
\end{equation}
The sampled datasets remain fixed within each game. For every sampled permutation \(\pi\), we sequentially form coalitions and compute
\begin{equation}
\Delta_i^{(g,t)}
=
v(S\cup\{i\}\mid D_{S\cup\{i\}}^{(g)})
-
v(S\mid D_S^{(g)}),
\label{eq:marginal}
\end{equation}
\revise{where \(D_S^{(g)}=\bigcup_{j\in S}D_j^{(g)}\) denotes the realized dataset of coalition \(S\).}

Averaging over all games and permutations yields
\begin{equation}
\widehat{\mathbb{E}[\phi_i]}
=
\frac{1}{n_{\text{games}}n_{\text{iter}}}
\sum_{g=1}^{n_{\text{games}}}
\sum_{t=1}^{n_{\text{iter}}}
\Delta_i^{(g,t)}.
\label{eq:estexp}
\end{equation}

\revise{With independent games, the variance is estimated by}
\begin{equation}
\widehat{\mathrm{Var}(\phi_i)}
=
\frac{1}{n_{\text{games}}-1}
\sum_{g=1}^{n_{\text{games}}}
\left(
\bar{\phi}_i^{(g)}
-
\widehat{\mathbb{E}[\phi_i]}
\right)^2,
\label{eq:estvar}
\end{equation}
\revise{where
\(
\bar{\phi}_i^{(g)}
=
n_{\text{iter}}^{-1}
\sum_{t=1}^{n_{\text{iter}}}
\Delta_i^{(g,t)}.
\)
If each realized-game Shapley value is computed exactly, Eq.~\eqref{eq:estvar} is unbiased for \(\mathrm{Var}(\phi_i)\); with finite \(n_{\text{iter}}\), it is unbiased for \(\mathrm{Var}(\bar{\phi}_i^{(g)})\), which additionally includes permutation-sampling error.}

\begin{revisionblock}
\begin{example}[Baseline estimation]
Consider three players, \(n_{\text{games}}=2\), and \(n_{\text{iter}}=2\). In game~1, permutations \((1,2,3)\) and \((2,3,1)\) produce coalition utilities \((0,2,5,9)\) and \((0,1,6,9)\), yielding marginal-contribution vectors \((2,3,4)\) and \((3,1,5)\). In game~2, permutations \((1,3,2)\) and \((3,2,1)\) produce coalition utilities \((0,1,4,8)\) and \((0,2,5,8)\), yielding marginal-contribution vectors \((1,4,3)\) and \((3,3,2)\).

Averaging over permutations gives
$
\bar{\boldsymbol{\phi}}^{(1)}=(2.5,2.0,4.5)$ and $
\bar{\boldsymbol{\phi}}^{(2)}=(2.0,3.5,2.5)$.
The estimated expected Shapley value is
\(
\widehat{\mathbb{E}[\boldsymbol{\phi}]}
=
\frac{1}{2}
\left(
\bar{\boldsymbol{\phi}}^{(1)}
+
\bar{\boldsymbol{\phi}}^{(2)}
\right)
=
(2.25,2.75,3.50)
\).
Applying Eq.~\eqref{eq:estvar} gives
\(
\widehat{\mathrm{Var}(\boldsymbol{\phi})}
=
(0.125,1.125,2.000).
\)
Thus, the inner average estimates the conditional Shapley value for each realized game, while the outer average estimates its expectation and variance across data realizations.
\end{example}
\end{revisionblock}

\revise{Although the baseline estimator is unbiased for the expected Shapley value, it requires \(O(n\,n_{\text{sample}}\,n_{\text{games}})\) fresh samples, making it expensive when source access is costly. The pooled methods reduce fresh-source access through sample reuse at the expense of finite-pool approximation and additional local computation.}

\begin{algorithm}[!t]
\caption{Baseline Method for Shapley Value Estimation}
\label{alg:baseline}
\begin{algorithmic}[1]
\REQUIRE Player distributions $\{P_i\}_{i=1}^n$, sample size $n_{\text{sample}}$, number of games $n_{\text{games}}$, iterations per game $n_{\text{iter}}$
\ENSURE Estimated Shapley values $\{\phi_i\}_{i=1}^n$
\STATE Initialize $\phi_i \gets 0$ for all $i$
\FOR{$g \gets 1$ to $n_{\text{games}}$}
    \FOR{$i \gets 1$ to $n$}
        \STATE Sample data $D_i^{(g)} \gets \{(x_j,y_j)\sim P_i\}_{j=1}^{n_{\text{sample}}}$
    \ENDFOR
    \STATE $\{\bar{\phi}_i^{(g)}\}_{i=1}^n \gets \text{\textsc{MC-Shapley}}(\{D_i^{(g)}\}_{i=1}^n,n_{\text{iter}})$
    \FOR{$i \gets 1$ to $n$}
        \STATE $\phi_i \gets \phi_i+\bar{\phi}_i^{(g)}$
    \ENDFOR
\ENDFOR
\RETURN $\{\phi_i/n_{\text{games}}\}_{i=1}^n$
\end{algorithmic}
\end{algorithm}

\section{Closed Form Solutions}\label{sec:theory}

In this section, we develop closed-form expressions for the expectation and variance of the Shapley value under probabilistic data distributions. \revise{Randomly sampled player data make coalition utilities and marginal contributions random variables, whereas each realized game defines a deterministic Shapley value.} We characterize the expectation and variance of \(\phi_i\), providing the analytical foundation for uncertainty-aware Shapley valuation.

\revise{Our analysis is related to, but distinct from, classical sampling-based Shapley approximation. Castro et al.~\cite{castro2009random} approximate the Shapley value of a fixed cooperative game by sampling player permutations, and subsequent work has developed error bounds and variance-reduction techniques~\cite{maleki2013bounding,castro2017improving}. In our setting, sampling player data induces a distribution over realized games. For each realization, the Shapley value is deterministic and can be computed exactly or approximated by these methods. Our focus is instead on characterizing its expectation and variance across data realizations.}

\subsection{Expectation of Shapley Value}

Let \(\phi_i\) denote the random Shapley value of player \(i\) induced by a realization of all players' data. We first characterize its expectation, \(\mathbb{E}[\phi_i]\).

\begin{proposition}
If \(\mathbb{E}[u(S\cup\{i\})-u(S)]\) admits a closed-form expression for every \(S\subseteq N\setminus\{i\}\), then so does \(\mathbb{E}[\phi_i]\).
\end{proposition}

\begin{proof}
Taking the expectation of Eq.~\eqref{eq:shapley_deterministic} and applying linearity of expectation,
\begin{align*}
\mathbb{E}[\phi_i]
=
\sum_{S\subseteq N\setminus\{i\}}
\frac{|S|!(n-|S|-1)!}{n!}
\,
\mathbb{E}\!\left[u(S\cup\{i\})-u(S)\right].
\end{align*}
Since the Shapley weights are constant, the expression is closed form whenever the expected marginal contributions are.
\end{proof}

\begin{example}\label{ex:closed-form-exp}
Suppose \(X_i\sim\mathcal{N}(\mu_i,\sigma_i^2)\) and
\(
u(S)=\sum_{k\in S}w_kX_k,
\)
where \(w_k\) is fixed. Then
\(
u(S\cup\{i\})-u(S)=w_iX_i,
\)
so
\(
\mathbb{E}[u(S\cup\{i\})-u(S)]=w_i\mu_i.
\)
Substituting into Eq.~\eqref{eq:shapley_deterministic} and using the fact that the Shapley weights sum to \(1\) gives
\(
\mathbb{E}[\phi_i]=w_i\mu_i.
\)
\end{example}

\subsection{Variance of Shapley Value}

The variance \(\mathrm{Var}(\phi_i)\) quantifies the uncertainty of player \(i\)'s contribution across data realizations.

\begin{proposition}\label{prop:closed_var}
If the variances and covariances of the marginal contributions \(u(S\cup\{i\})-u(S)\) admit closed-form expressions for all \(S,T\subseteq N\setminus\{i\}\), then so does \(\mathrm{Var}(\phi_i)\).
\end{proposition}

\begin{proof}
From Eq.~\eqref{eq:shapley_deterministic},
\begin{align*}
\mathrm{Var}(\phi_i)
&= \mathrm{Var}\!\left( \sum_{S} c_S \, (u(S \cup \{i\}) - u(S)) \right) \\
&= \sum_{S} c_S^2 \, \mathrm{Var}(u(S \cup \{i\}) - u(S)) \\
&\quad + \sum_{S \neq T} c_S c_T \,
    \mathrm{Cov}(u(S \cup \{i\}) - u(S), \\
&\qquad\qquad\qquad\quad
    u(T \cup \{i\}) - u(T)).
\end{align*}
where
\(
c_S=\frac{|S|!(n-|S|-1)!}{n!}
\)
and
\(
c_T=\frac{|T|!(n-|T|-1)!}{n!}.
\)
Since the coefficients are constant, \(\mathrm{Var}(\phi_i)\) is closed form whenever all variances and covariances of the marginal contributions are.
\end{proof}

\begin{example}
For the weighted additive Gaussian model in Example~\ref{ex:closed-form-exp},
\(
u(S\cup\{i\})-u(S)=w_iX_i
\)
for every
\(S\subseteq N\setminus\{i\}\).
Since the Shapley weights sum to \(1\),
\(
\phi_i=w_iX_i,
\)
and therefore
\[
\mathrm{Var}(\phi_i)=w_i^2\sigma_i^2.
\]
\end{example}

\subsection{Estimated Expectation of Shapley Value}

We now analyze the empirical estimator of the expected Shapley value. Let \(m\) denote the number of independently sampled games. For each game \(j\), let \(\phi_i^{(j)}\) be the Shapley value of player \(i\). The estimator is
\begin{equation}
\widehat{\mathbb{E}[\phi_i]} = \frac{1}{m} \sum_{j=1}^{m} \phi_i^{(j)}.
\label{eq:est_exp}
\end{equation}

\begin{proposition}
The estimator in Eq.~\eqref{eq:est_exp} is unbiased, i.e., \\
\(\mathbb{E}[\widehat{\mathbb{E}\allowbreak[\phi_i]}]\allowbreak = \mathbb{E}\allowbreak[\phi_i]\).
\end{proposition}

\begin{proof}
Since each \(\phi_i^{(j)}\) is an i.i.d. realization of \(\phi_i\),
$
\mathbb{E}[\widehat{\mathbb{E}[\phi_i]}]
= \frac{1}{m} \sum_{j=1}^m \mathbb{E}[\phi_i^{(j)}]
= \mathbb{E}[\phi_i].
$
\end{proof}

\begin{proposition} 
If \(\mathrm{Var}(\phi_i)<\infty\), then the variance of the estimator \(\widehat{\mathbb{E}[\phi_i]}\) is
\begin{equation}
\mathrm{Var}\!\left(\widehat{\mathbb{E}[\phi_i]}\right)
= \frac{1}{m} \, \mathrm{Var}(\phi_i).
\label{eq:var_est_exp}
\end{equation}
\end{proposition}

\begin{proof}
Using independence of \(\phi_i^{(j)}\),
$
\mathrm{Var}\!\left(\widehat{\mathbb{E}[\phi_i]}\right)
= \mathrm{Var}\!\left(\frac{1}{m} \sum_{j=1}^m \phi_i^{(j)}\right)
= \frac{1}{m^2} \sum_{j=1}^m \mathrm{Var}(\phi_i)
= \frac{1}{m} \, \mathrm{Var}(\phi_i)$.
\end{proof}

\subsection{Estimated Variance of Shapley Value}

The sample variance over \(m\) realizations is
\begin{equation}
\widehat{\mathrm{Var}(\phi_i)}
=
\frac{1}{m-1}
\sum_{j=1}^{m}
\left(
\phi_i^{(j)}
-
\widehat{\mathbb{E}[\phi_i]}
\right)^2.
\label{eq:est_var}
\end{equation}

\begin{proposition}
The estimator in Eq.~\eqref{eq:est_var} is unbiased:
\[
\mathbb{E}[\widehat{\mathrm{Var}(\phi_i)}]
=
\mathrm{Var}(\phi_i).
\]
\end{proposition}

\begin{proof}
Expanding Eq.~\eqref{eq:est_var} and applying the variance decomposition identity,
\[
\mathbb{E}[\widehat{\mathrm{Var}(\phi_i)}]
=
\frac{1}{m-1}
\left(
m\,\mathrm{Var}(\phi_i)
-
m\,\mathrm{Var}\!\left(\widehat{\mathbb{E}[\phi_i]}\right)
\right).
\]
Substituting Eq.~\eqref{eq:var_est_exp} gives
\[
\mathbb{E}[\widehat{\mathrm{Var}(\phi_i)}]
=
\frac{m-1}{m-1}
\mathrm{Var}(\phi_i)
=
\mathrm{Var}(\phi_i).
\]
\end{proof}

The variance of this estimator is~\cite{Mood1974}
\begin{equation}
\mathrm{Var}\!\left(\widehat{\mathrm{Var}(\phi_i)}\right)
=
\frac{1}{m}
\left(
\mu_4
-
\frac{m-3}{m-1}\sigma^4
\right),
\label{eq:var_var}
\end{equation}
where
\(
\mu_4=\mathbb{E}\!\left[(\phi_i-\mathbb{E}[\phi_i])^4\right]
\)
is the fourth central moment and
\(
\sigma^4=(\mathrm{Var}(\phi_i))^2.
\)
\revise{For general utility functions, \(\mu_4\) may not admit a tractable closed-form expression. Consequently, neither may the variance in Eq.~\eqref{eq:var_var} without additional distributional assumptions.}
Nevertheless, Eq.~\eqref{eq:var_var} gives the exact finite-sample variance of the estimator.

\section{Empirical Estimation Approaches} 
\label{sec:empirical}

Closed-form analysis is possible only when the expected marginal contributions admit closed-form expressions. In general, however, player data distributions do not satisfy this condition. This section therefore develops empirical estimation methods for the general setting.

\subsection{Experimental Setup}

We first describe the synthetic and Wine Quality experiments used throughout the paper. The Digits, sensitivity, scalability, and runtime experiments are introduced with their corresponding results.

\subsubsection{Dataset Preparation}

The core experiments use a synthetic regression task and the Wine Quality regression dataset.

\textbf{Synthetic Dataset:}
The synthetic dataset is generated from
\(f(x)=p(x)+s(x)\), where
\(p(x)=\sum_{k=0}^{K}c_kx^k\) with
\(c_0=0.0\), \(c_1=0.3\), and \(c_2=-0.02\), and
\(s(x)=A\sin(2\pi\nu x+\phi)+b\) with
\(A=2.0\), \(\nu=0.5\), \(\phi=\pi/4\), and \(b=3.0\).
We generate \(n=10\) players, each with a Gaussian distribution
\(\mathcal{N}(\mu_i,\sigma_i^2)\), where
\(\mu_i\) is uniformly spaced over \([-7,7]\) and
\(\sigma_i\) is uniformly sampled from \([0.5,2.5]\).
Each player samples \(x\)-values, clipped to \([-10,10]\), and assigns labels
\(y=f(x)\).
Model accuracy is evaluated on 1,000 uniformly spaced validation points. This construction gives different players distinct regions of the underlying function.

\textbf{Real Dataset:}
We use the Wine Quality dataset~\cite{cortez2009modeling}, which contains 4,898 white-wine samples with 11 physicochemical attributes and quality scores from 3 to 9.
To simulate heterogeneous players, we partition the training data by the \texttt{alcohol} feature. After shuffling with seed 42, we split the data into 70\% training (3,429 samples) and 30\% validation (1,469 samples). Let
\(\{q_0,\ldots,q_{10}\}\)
be the alcohol deciles in the training set. Player \(i\) receives samples with alcohol values in
\([q_{i-1},q_i)\), yielding ten players with 227--450 samples each. During each game, we subsample each player's partition, preserving distributional heterogeneity while introducing sampling variability.

\subsubsection{Utility Function}

For both datasets, the utility is the inverse mean squared error (MSE) on a fixed validation set.

\textbf{Synthetic Data Utility:}
For each game, player \(i\) provides a bootstrap sample
\(\tilde{\mathcal D}_i=\{(\tilde x_i^{(k)},\tilde y_i^{(k)})\}_{k=1}^{m}\).
For coalition \(S\),
\(
\tilde{\mathcal D}_S=\bigcup_{i\in S}\tilde{\mathcal D}_i.
\)
Predictions are obtained by nearest-neighbor interpolation,
\(
\hat y_{\mathrm{val}}=y,
(x,y)=\arg\min_{(x',y')\in\tilde{\mathcal D}_S}|x'-x_{\mathrm{val}}|,
\)
and the utility is
\(
v(S;\tilde{\mathcal D}_S)=\frac{1}{1+\mathrm{MSE}(S)},
\)
where
$
\mathrm{MSE}(S)
=
\frac{1}{|\mathcal V|}
\sum_{(x_{\mathrm{val}},y_{\mathrm{val}})\in\mathcal V}
(\hat y_{\mathrm{val}}-y_{\mathrm{val}})^2.
$
Different bootstrap samples of the same coalition therefore produce different utilities.

\textbf{Real Data Utility:}
For the Wine Quality data, player \(j\) owns index set
\(\mathcal I_j\).
In each game, we sample
\(\tilde{\mathcal I}_j\subset\mathcal I_j\)
without replacement and form
$
\tilde{\mathcal D}_S
=
\{
(\mathbf X_i,y_i):
i\in\bigcup_{j\in S}\tilde{\mathcal I}_j
\}$.
A linear regression model is trained on
\(\tilde{\mathcal D}_S\),
with utility
\(
v(S;\tilde{\mathcal D}_S)
=
\frac{1}{1+\mathrm{MSE}(S)},
\)
where
\(
\mathrm{MSE}(S)
=
\frac{1}{|\mathcal V|}
\sum_{t=1}^{|\mathcal V|}
(\hat y_{\mathrm{val}}^{(t)}-y_{\mathrm{val}}^{(t)})^2,
\)
\(
\hat y_{\mathrm{val}}^{(t)}
=
\boldsymbol\beta_S^\top
\mathbf X_{\mathrm{val}}^{(t)},
\)
and
\(\boldsymbol\beta_S\)
is obtained by ordinary least squares.
Coalitions with fewer than two samples are assigned zero utility.

\subsubsection{Evaluation Metrics}

\revise{To evaluate estimation quality, we perform \(R\) independent replications, each containing \(G\) games. The baseline independently samples every game. The pooled methods construct fresh pools once per replication and generate games by independently bootstrapping from the realized pools. Thus, games are conditionally independent given the pools but may be marginally correlated.}

For player \(i\), replication \(r\) produces
\(\{\phi_i^{(r,g)}\}_{g=1}^{G}\).
Within each replication,
\begin{align}
\widehat{\mathbb E}_r[\phi_i]
&=
\frac1G
\sum_{g=1}^{G}
\phi_i^{(r,g)},
\\
\widehat{\mathrm{Var}}_r(\phi_i)
&=
\frac1{G-1}
\sum_{g=1}^{G}
\left(
\phi_i^{(r,g)}
-
\widehat{\mathbb E}_r[\phi_i]
\right)^2.
\end{align}
Across replications,
\begin{align}
\mathrm{Var}(\widehat{\mathbb E}[\phi_i])
&=
\frac1{R-1}
\sum_{r=1}^{R}
\left(
\widehat{\mathbb E}_r[\phi_i]
-
\frac1R
\sum_{r'}
\widehat{\mathbb E}_{r'}[\phi_i]
\right)^2,
\\
\mathrm{Var}(\widehat{\mathrm{Var}}(\phi_i))
&=
\frac1{R-1}
\sum_{r=1}^{R}
\left(
\widehat{\mathrm{Var}}_r(\phi_i)
-
\frac1R
\sum_{r'}
\widehat{\mathrm{Var}}_{r'}(\phi_i)
\right)^2.
\end{align}
For the ten-player experiments,
\begin{align}
\overline{\mathrm{Var}(\widehat{\mathbb E}[\phi])}
&=
\frac1n
\sum_{i=1}^{n}
\mathrm{Var}(\widehat{\mathbb E}[\phi_i]),
\label{eq:avg_var_mean}
\\
\overline{\mathrm{Var}(\widehat{\mathrm{Var}}(\phi))}
&=
\frac1n
\sum_{i=1}^{n}
\mathrm{Var}(\widehat{\mathrm{Var}}(\phi_i)).
\label{eq:avg_var_variance}
\end{align}

The first metric measures the stability of the estimated expectation, and the second that of the estimated variance. Lower values indicate more stable estimation.

\subsection{Baseline Method: Sources of Uncertainty}

We first analyze the baseline method to quantify the effects of two uncertainty sources: (i) sampling uncertainty from independently sampled games and (ii) computation uncertainty from finite Monte Carlo permutations. We study their effects by varying \(n_{\text{games}}\) and \(n_{\text{iter}}\).

\begin{figure}[!t]
\centering
\includegraphics[width=\linewidth]{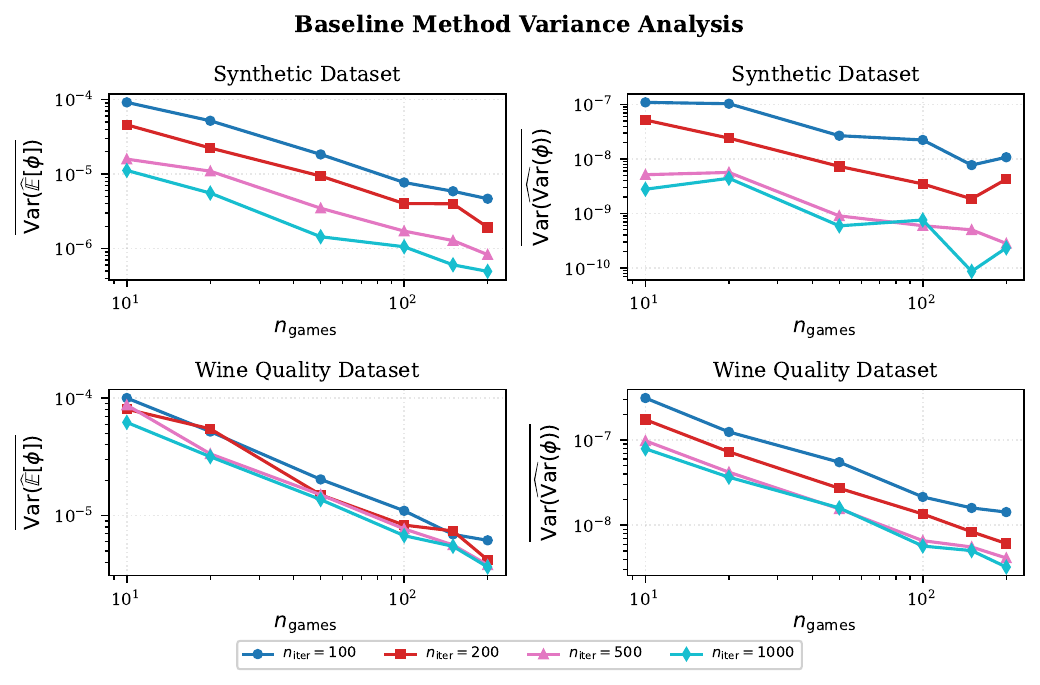}
\caption{Effect of $n_{\text{games}}$ and $n_{\text{iter}}$ on variance metrics for the baseline method. 
Increasing either parameter reduces both variance metrics across datasets, with $n_{\text{games}}$ dominating mean stability and $n_{\text{iter}}$ dominating variance stability.}
\label{fig:variance:baseline}
\end{figure}

Table~\ref{tab:variance_naive_combined} and Figure~\ref{fig:variance:baseline} report the average variance of the estimated expectation, \(\overline{\mathrm{Var}(\widehat{\mathbb{E}}[\phi])}\), and variance, \(\overline{\mathrm{Var}(\widehat{\mathrm{Var}}(\phi))}\), over all players. Lower values indicate more stable estimation.

\subsubsection{Impact of Sampling Uncertainty}

Increasing \(n_{\text{games}}\) consistently reduces both metrics. With \(n_{\text{iter}}=100\), \(\overline{\mathrm{Var}(\widehat{\mathbb{E}}[\phi])}\) decreases by 94.9\% on the synthetic dataset and 93.9\% on Wine Quality as \(n_{\text{games}}\) increases from 10 to 200. Across all settings, the variance decreases approximately as \(1/n_{\text{games}}\), confirming that repeated sampling effectively suppresses sampling uncertainty.

The same trend holds for \(\overline{\mathrm{Var}(\widehat{\mathrm{Var}}(\phi))}\), which decreases by about one order of magnitude over the same range of \(n_{\text{games}}\). These results identify sampling uncertainty as the dominant source of estimation variance.

\subsubsection{Impact of Computation Uncertainty}

Increasing \(n_{\text{iter}}\) also improves estimation, particularly for variance estimation. For \(\overline{\mathrm{Var}(\widehat{\mathbb{E}}[\phi])}\), increasing \(n_{\text{iter}}\) from 100 to 1000 reduces the variance by 40--90\%, with larger gains at higher \(n_{\text{games}}\). The effect is stronger for \(\overline{\mathrm{Var}(\widehat{\mathrm{Var}}(\phi))}\), where the reduction reaches 70--99\% across all configurations. Thus, Monte Carlo averaging is particularly important for accurately estimating the Shapley-value variance.

\subsubsection{Summary}

Sampling and computation uncertainty affect different aspects of estimation. Increasing \(n_{\text{games}}\) primarily improves the stability of the estimated expectation, whereas increasing \(n_{\text{iter}}\) is more effective for stabilizing the estimated variance. These observations motivate the sampling-efficient methods developed next.

\begin{table*}[!t]
\centering
\caption{Variance metrics for the baseline method across synthetic and real datasets. Both \(\overline{\mathrm{Var}(\widehat{\mathbb{E}}[\phi])}\) and \(\overline{\mathrm{Var}(\widehat{\mathrm{Var}}(\phi))}\) generally decrease as \(n_{\text{games}}\) and \(n_{\text{iter}}\) increase, confirming that sampling and computation uncertainty both affect Shapley value estimation.}
\label{tab:variance_naive_combined}
\resizebox{0.75\textwidth}{!}{%
\begin{tabular}{
l
S[table-format=1.2e-1]
S[table-format=1.2e-1]
S[table-format=1.2e-1]
S[table-format=1.2e-1]
S[table-format=1.2e-1]
S[table-format=1.2e-1]
S[table-format=1.2e-1]
S[table-format=1.2e-1]
}
\toprule
\multicolumn{9}{c}{Baseline Method \(\overline{\mathrm{Var}(\widehat{\mathbb{E}}[\phi])}\)} \\
\midrule
{\(n_{\text{games}}\)} & \multicolumn{4}{c}{Synthetic} & \multicolumn{4}{c}{Real (Wine Quality)} \\
\cmidrule(lr){2-5}\cmidrule(lr){6-9}
& {\(n_{\text{iter}}=100\)}
& {\(n_{\text{iter}}=200\)}
& {\(n_{\text{iter}}=500\)}
& {\(n_{\text{iter}}=1000\)}
& {\(n_{\text{iter}}=100\)}
& {\(n_{\text{iter}}=200\)}
& {\(n_{\text{iter}}=500\)}
& {\(n_{\text{iter}}=1000\)} \\
\midrule
10  & 9.16e-05 & 4.57e-05 & 1.59e-05 & 1.12e-05 & 1.00e-04 & 8.05e-05 & 8.73e-05 & 6.22e-05 \\
20  & 5.19e-05 & 2.24e-05 & 1.09e-05 & 5.56e-06 & 5.19e-05 & 5.48e-05 & 3.35e-05 & 3.17e-05 \\
50  & 1.83e-05 & 9.40e-06 & 3.48e-06 & 1.44e-06 & 2.03e-05 & 1.50e-05 & 1.50e-05 & 1.37e-05 \\
100 & 7.70e-06 & 4.02e-06 & 1.72e-06 & 1.06e-06 & 1.10e-05 & 8.29e-06 & 7.70e-06 & 6.74e-06 \\
150 & 5.88e-06 & 3.99e-06 & 1.28e-06 & 6.05e-07 & 6.91e-06 & 7.39e-06 & 5.62e-06 & 5.47e-06 \\
200 & 4.67e-06 & 1.93e-06 & 8.28e-07 & 4.94e-07 & 6.15e-06 & 4.19e-06 & 3.80e-06 & 3.64e-06 \\
\midrule
\multicolumn{9}{c}{Baseline Method \(\overline{\mathrm{Var}(\widehat{\mathrm{Var}}(\phi))}\)} \\
\midrule
{\(n_{\text{games}}\)} & \multicolumn{4}{c}{Synthetic} & \multicolumn{4}{c}{Real (Wine Quality)} \\
\cmidrule(lr){2-5}\cmidrule(lr){6-9}
& {\(n_{\text{iter}}=100\)}
& {\(n_{\text{iter}}=200\)}
& {\(n_{\text{iter}}=500\)}
& {\(n_{\text{iter}}=1000\)}
& {\(n_{\text{iter}}=100\)}
& {\(n_{\text{iter}}=200\)}
& {\(n_{\text{iter}}=500\)}
& {\(n_{\text{iter}}=1000\)} \\
\midrule
10  & 1.10e-07 & 5.20e-08 & 5.15e-09 & 2.78e-09 & 3.11e-07 & 1.73e-07 & 9.79e-08 & 7.90e-08 \\
20  & 1.04e-07 & 2.42e-08 & 5.67e-09 & 4.43e-09 & 1.24e-07 & 7.27e-08 & 4.20e-08 & 3.66e-08 \\
50  & 2.68e-08 & 7.32e-09 & 9.03e-10 & 5.92e-10 & 5.50e-08 & 2.72e-08 & 1.55e-08 & 1.60e-08 \\
100 & 2.24e-08 & 3.48e-09 & 5.96e-10 & 7.58e-10 & 2.15e-08 & 1.35e-08 & 6.59e-09 & 5.70e-09 \\
150 & 7.73e-09 & 1.85e-09 & 5.00e-10 & 8.62e-11 & 1.60e-08 & 8.38e-09 & 5.56e-09 & 5.04e-09 \\
200 & 1.08e-08 & 4.26e-09 & 2.81e-10 & 2.33e-10 & 1.43e-08 & 6.10e-09 & 4.12e-09 & 3.23e-09 \\
\bottomrule
\end{tabular}%
}
\end{table*}

\subsection{Pooled Method: Efficient Sampling Through Reuse}

The baseline method draws fresh samples from each player's distribution in every game, which can be expensive when source access is limited. We therefore propose the \textbf{pooled method}, which constructs one reusable sample pool for each player and generates games by bootstrap resampling from the realized pools.
\revise{Bootstrap resampling itself is standard; our contribution is its use as a source-access layer for probabilistic Shapley estimation, together with an explicit distinction between the population, finite-pool, and conditional bootstrap targets.}

\subsubsection{Method Description}

Each player \(i\) first draws \(n_{\text{pool}}\) samples from \(P_i\) to construct a reusable pool
\(
\mathcal{S}_i=\{(x_j,y_j)\}_{j=1}^{n_{\text{pool}}}.
\)
For each game \(g\), player \(i\) then draws a bootstrap sample
\(
\mathcal{B}_i^{(g)}
\)
of size \(n_{\text{boot}}\) from \(\mathcal{S}_i\). For coalition \(S\),
\(
\mathcal{B}_S^{(g)}
=
\bigcup_{i\in S}
\mathcal{B}_i^{(g)},
\)
and the marginal contribution is
\[
\Delta_i^{(g,t)}
=
v(S\cup\{i\}\mid\mathcal{B}_{S\cup\{i\}}^{(g)})
-
v(S\mid\mathcal{B}_S^{(g)}).
\]
The pooled method reduces fresh-source sampling from
\(O(n_{\text{games}}n_{\text{boot}}n)\)
to
\(O(n_{\text{pool}}n)\)
because each pool is constructed only once.

\begin{revisionblock}
\subsubsection{Estimation Targets and Pool-Induced Dependence}

Let \(\Phi_i(D)\) denote player \(i\)'s exact Shapley value for independently sampled player datasets \(D\). The population targets are
\(
\mu_i=\mathbb{E}_D[\Phi_i(D)]
\)
and
\(
V_i=\mathrm{Var}_D(\Phi_i(D)).
\)
A pooled replication instead observes finite pools \(\mathcal S_r\). Let
\(Y_i^{(r,g)}\)
denote the estimate from bootstrap game \(g\), including permutation-sampling error when \(n_{\text{iter}}\) is finite. Conditional on \(\mathcal S_r\), define
\begin{equation}
\mu_{i,\mathcal S_r}
=
\mathbb E[Y_i^{(r,g)}\mid\mathcal S_r],
\qquad
V_{i,\mathcal S_r}
=
\mathrm{Var}(Y_i^{(r,g)}\mid\mathcal S_r).
\label{eq:conditional-pool-targets}
\end{equation}

\begin{proposition}
Suppose each game uses independent bootstrap, permutation, and algorithmic randomness. Conditional on \(\mathcal S_r\), the pooled games are i.i.d. Hence, their sample mean and sample variance are unbiased for \(\mu_{i,\mathcal S_r}\) and \(V_{i,\mathcal S_r}\), respectively. Sharing a random pool does not bias these conditional estimators, although the conditional bootstrap targets need not equal the population targets \((\mu_i,V_i)\).
\end{proposition}

\begin{proof}
Conditional on \(\mathcal S_r\), the variables
\(Y_i^{(r,1)},\ldots,Y_i^{(r,G)}\)
are i.i.d. Standard sample-mean and Bessel-corrected sample-variance identities therefore give
\begin{equation}
\mathbb E[\widehat{\mathbb E}_r[\phi_i]\mid\mathcal S_r]
=
\mu_{i,\mathcal S_r},
\qquad
\mathbb E[\widehat{\mathrm{Var}}_r(\phi_i)\mid\mathcal S_r]
=
V_{i,\mathcal S_r}.
\label{eq:conditional-pool-unbiased}
\end{equation}
For \(g\neq h\), the law of total covariance gives
\begin{equation}
\mathrm{Cov}(Y_i^{(r,g)},Y_i^{(r,h)})
=
\mathrm{Var}(\mu_{i,\mathcal S_r}),
\label{eq:pooled-game-covariance}
\end{equation}
and the law of total variance gives
\begin{equation}
\mathrm{Var}(Y_i^{(r,g)})
=
\mathbb E[V_{i,\mathcal S_r}]
+
\mathrm{Var}(\mu_{i,\mathcal S_r}).
\label{eq:pooled-total-variance}
\end{equation}
Thus, the within-replication sample variance estimates the conditional bootstrap variance rather than the unconditional variance, omitting the between-pool component. The difference from \(V_i\) is a finite-pool approximation error that vanishes only when the bootstrap is consistent as \(n_{\text{pool}}\) increases. Finite \(n_{\text{iter}}\) additionally introduces Monte Carlo error.
\end{proof}

Because each replication constructs independent pools,
\(
\widehat{\mathbb E}_r[\phi_i]
\)
and
\(
\widehat{\mathrm{Var}}_r(\phi_i)
\)
are i.i.d. across replications. Consequently, the sample variances in Eqs.~\eqref{eq:avg_var_mean} and~\eqref{eq:avg_var_variance} are unbiased for the corresponding across-replication variances. These metrics quantify the stability of the complete estimation procedure, including pool construction, and therefore retain finite-pool and finite-permutation approximation error.
\end{revisionblock}

\begin{algorithm}[!t]
\caption{Pooled Method for Shapley Value Estimation}
\label{alg:pooled}
\begin{algorithmic}[1]
\REQUIRE Player distributions $\{P_i\}_{i=1}^n$, pool size $n_{\text{pool}}$, bootstrap size $n_{\text{boot}}$, number of games $n_{\text{games}}$, iterations per game $n_{\text{iter}}$
\ENSURE Estimated Shapley values $\{\phi_i\}_{i=1}^n$
\STATE Initialize $\phi_i \gets 0$ for all $i$
\FOR{$i \gets 1$ to $n$}
    \STATE Sample pool $\mathcal{S}_i \gets \{(x_j,y_j)\sim P_i\}_{j=1}^{n_{\text{pool}}}$
\ENDFOR
\FOR{$g \gets 1$ to $n_{\text{games}}$}
    \FOR{$i \gets 1$ to $n$}
        \STATE Bootstrap $\mathcal{B}_i^{(g)} \gets \text{sample}(\mathcal{S}_i,n_{\text{boot}},\text{replace}=\text{True})$
    \ENDFOR
    \STATE $\{\bar{\phi}_i^{(g)}\}_{i=1}^n \gets \text{\textsc{MC-Shapley}}(\{\mathcal{B}_i^{(g)}\}_{i=1}^n,n_{\text{iter}})$
    \FOR{$i \gets 1$ to $n$}
        \STATE $\phi_i \gets \phi_i+\bar{\phi}_i^{(g)}$
    \ENDFOR
\ENDFOR
\RETURN $\{\phi_i/n_{\text{games}}\}_{i=1}^n$
\end{algorithmic}
\end{algorithm}

\revise{
\subsubsection{Sampling Efficiency}

We count a \emph{fresh source sample} only when it is drawn directly from a player's distribution; bootstrap samples from cached pools are excluded. The baseline requires \(n\,n_{\text{games}}\,n_{\text{sample}}\) fresh source samples, whereas each pooled estimator requires only \(n\,n_{\text{pool}}\) samples for one-time pool construction. For example, with \(n=10\), \(n_{\text{games}}=200\), \(n_{\text{sample}}=n_{\text{boot}}=1000\), and \(n_{\text{pool}}=5000\), pooling reduces fresh source samples from \(2\times10^6\) to \(5\times10^4\) (97.5\%). Pool construction is amortized across all games, while subsequent bootstrap sampling is performed locally. Fresh-source counts measure data-access cost rather than wall-clock time, which is evaluated separately in Section~\ref{sec:measured-efficiency}.
}

\begin{revisionblock}
\subsubsection{Comparison with the Baseline at a Matched Evaluation Budget}

We compare the baseline and pooled methods using the same \(n_{\text{games}}\), \(n_{\text{iter}}\), and per-game sample size (\(n_{\text{sample}}=n_{\text{boot}}\)). The pooled configurations are \((n_{\text{pool}},n_{\text{boot}})=(5000,1000)\) for the synthetic data and \((250,60)\) for Wine Quality. Thus, both methods perform the same number of utility evaluations, while the baseline draws fresh source samples for every game and the pooled method reuses a single realized pool.

\begin{figure}[!t]
\centering
\includegraphics[width=\linewidth]{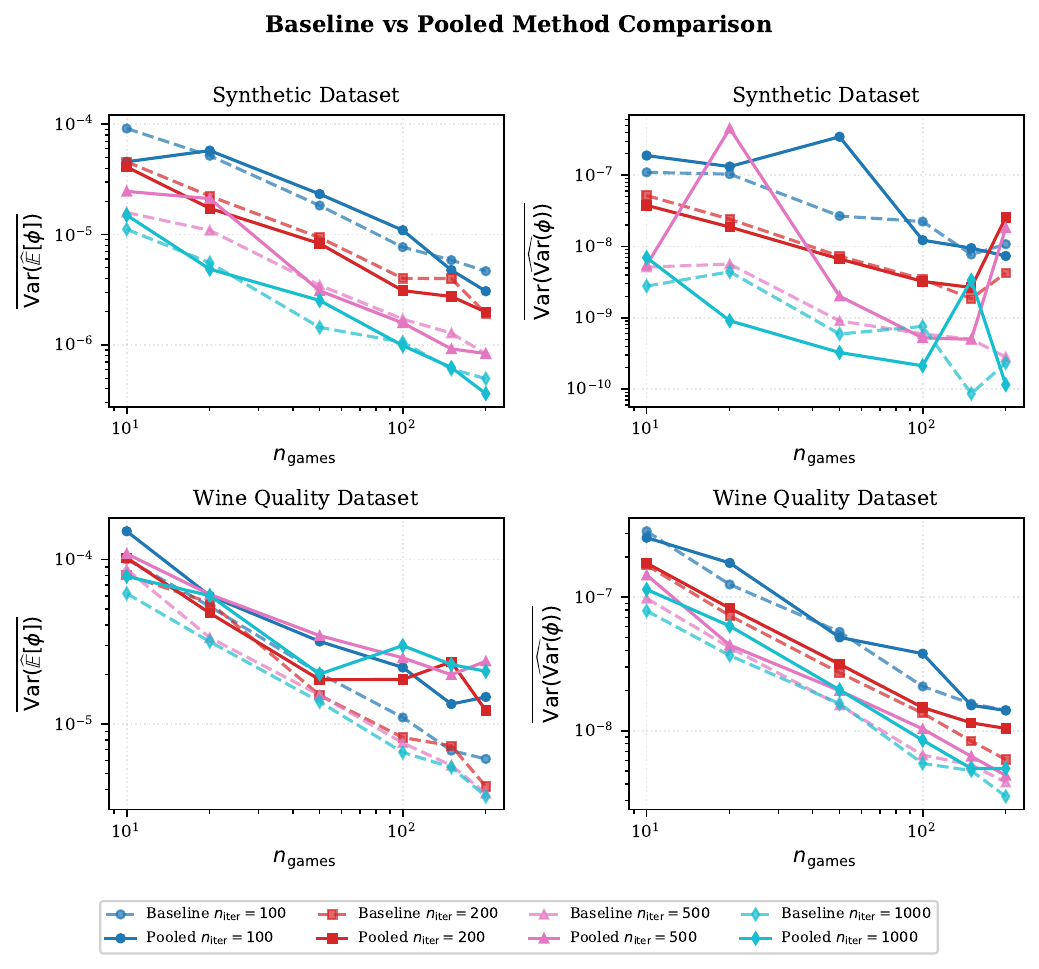}
\caption{Comparison of the baseline and pooled methods at matched numbers of games and Monte Carlo iterations. Both variance metrics generally decrease as either budget increases, but neither method uniformly dominates under a fixed evaluation budget.}
\label{fig:baseline_vs_pooled}
\end{figure}

Figure~\ref{fig:baseline_vs_pooled} shows that pooling preserves the convergence behavior of fresh sampling but does not uniformly reduce estimator variance under a fixed evaluation budget. The synthetic results are mixed, whereas the baseline generally achieves lower across-replication variance on Wine Quality. This behavior reflects the dependence introduced by reusing a finite pool: additional bootstrap games reduce conditional Monte Carlo variation but cannot eliminate between-pool variation. This experiment isolates the statistical effect of sample reuse under a fixed evaluation budget. Section~\ref{sec:source-budget-variance} instead fixes source access, allowing pooling to convert the saved source samples into additional local games.
\end{revisionblock}

\begin{table*}[!t]
\centering
\caption{Variance metrics for the pooled method across datasets (\(n_{\text{pool}}=5000\), \(n_{\text{boot}}=1000\) for synthetic; \(n_{\text{pool}}=250\), \(n_{\text{boot}}=60\) for real). Both metrics generally decrease as \(n_{\text{games}}\) and \(n_{\text{iter}}\) increase, showing that bootstrap reuse maintains estimation quality.}
\label{tab:variance_pool_combined_250_60}
\resizebox{0.8\textwidth}{!}{%
\begin{tabular}{
l
S[table-format=1.2e-1]
S[table-format=1.2e-1]
S[table-format=1.2e-1]
S[table-format=1.2e-1]
S[table-format=1.2e-1]
S[table-format=1.2e-1]
S[table-format=1.2e-1]
S[table-format=1.2e-1]
}
\toprule
\multicolumn{9}{c}{Pooled Method \(\overline{\mathrm{Var}(\widehat{\mathbb{E}}[\phi])}\)} \\
\midrule
{\(n_{\text{games}}\)} & \multicolumn{4}{c}{Synthetic} & \multicolumn{4}{c}{Real (Wine Quality)} \\
\cmidrule(lr){2-5}\cmidrule(lr){6-9}
& {\(n_{\text{iter}}=100\)}
& {\(n_{\text{iter}}=200\)}
& {\(n_{\text{iter}}=500\)}
& {\(n_{\text{iter}}=1000\)}
& {\(n_{\text{iter}}=100\)}
& {\(n_{\text{iter}}=200\)}
& {\(n_{\text{iter}}=500\)}
& {\(n_{\text{iter}}=1000\)} \\
\midrule
10  & 4.58e-05 & 4.11e-05 & 2.47e-05 & 1.50e-05 & 1.48e-04 & 1.02e-04 & 1.08e-04 & 7.89e-05 \\
50  & 2.33e-05 & 8.30e-06 & 3.10e-06 & 2.53e-06 & 3.18e-05 & 1.86e-05 & 3.45e-05 & 2.02e-05 \\
100 & 1.10e-05 & 3.11e-06 & 1.58e-06 & 9.88e-07 & 2.21e-05 & 1.87e-05 & 2.53e-05 & 3.00e-05 \\
200 & 3.07e-06 & 1.99e-06 & 8.36e-07 & 3.62e-07 & 1.46e-05 & 1.21e-05 & 2.41e-05 & 2.09e-05 \\
\midrule
\multicolumn{9}{c}{Pooled Method \(\overline{\mathrm{Var}(\widehat{\mathrm{Var}}(\phi))}\)} \\
\midrule
{\(n_{\text{games}}\)} & \multicolumn{4}{c}{Synthetic} & \multicolumn{4}{c}{Real (Wine Quality)} \\
\cmidrule(lr){2-5}\cmidrule(lr){6-9}
& {\(n_{\text{iter}}=100\)}
& {\(n_{\text{iter}}=200\)}
& {\(n_{\text{iter}}=500\)}
& {\(n_{\text{iter}}=1000\)}
& {\(n_{\text{iter}}=100\)}
& {\(n_{\text{iter}}=200\)}
& {\(n_{\text{iter}}=500\)}
& {\(n_{\text{iter}}=1000\)} \\
\midrule
10  & 1.89e-07 & 3.78e-08 & 5.49e-09 & 6.99e-09 & 2.78e-07 & 1.80e-07 & 1.46e-07 & 1.14e-07 \\
50  & 3.47e-07 & 6.69e-09 & 2.01e-09 & 3.26e-10 & 5.00e-08 & 3.15e-08 & 2.00e-08 & 2.02e-08 \\
100 & 1.23e-08 & 3.24e-09 & 5.23e-10 & 2.12e-10 & 3.78e-08 & 1.49e-08 & 1.03e-08 & 8.51e-09 \\
200 & 7.37e-09 & 2.57e-08 & 1.83e-08 & 1.15e-10 & 1.42e-08 & 1.04e-08 & 4.64e-09 & 5.23e-09 \\
\bottomrule
\end{tabular}%
}
\end{table*}

\subsection{Stratified Pooled Method: Variance-Adaptive Allocation}

The pooled method reduces source access through sample reuse but assigns every player the same bootstrap size. When player distributions have different dispersions, they contribute different amounts of sampling variability to coalition evaluation.
\revise{Players with more dispersed data can induce greater sampling variability than those with less dispersed data.}
This motivates allocating more local samples to higher-variance players.

We propose the \textbf{stratified pooled method}, which allocates player-specific bootstrap sizes according to empirical variance.
\revise{Motivated by variance-adaptive allocation in stratified sampling~\cite{neyman1992two},}
it redistributes a fixed bootstrap budget according to player variability. Unlike uniform pooling, it performs exact-budget allocation rather than variance scoring alone. The goal is to improve estimation without increasing the total number of bootstrap samples per game; it is not intended to dominate uniform pooling for every dataset or metric.

\subsubsection{Method Description}

\begin{revisionblock}

The stratified pooled method extends the pooled method in three phases.

\paragraph{Pool construction}
Each player \(i\) draws \(n_{\text{pool}}\) samples from \(P_i\) to form a reusable pool
\(
\mathcal{S}_i=\{(x_j^{(i)},y_j^{(i)})\}_{j=1}^{n_{\text{pool}}},
\)
as in the pooled method.

\paragraph{Variance analysis}
Let
\(
\mathcal{S}_{\mathrm{all}}
=
\operatorname{concat}
(\mathcal{S}_1,\ldots,\mathcal{S}_n)
\)
and
\(
\mathcal{K}
=
\{k:\operatorname{Var}(\mathcal{S}_{\mathrm{all}}^{(k)})>0\}.
\)
Player \(i\)'s variability is measured by
\(
\sigma_i^2
=
\sum_{k\in\mathcal K}
\frac{\operatorname{Var}(\mathcal S_i^{(k)})}
{\operatorname{Var}(\mathcal S_{\mathrm{all}}^{(k)})}.
\)
Using global feature variances normalizes heterogeneous feature scales while preserving differences among players. For nonconstant scores,
\(
\tilde{\sigma}_i^2
=
\frac{\sigma_i^2-\min_j\sigma_j^2}
{\max_j\sigma_j^2-\min_j\sigma_j^2}.
\)

\paragraph{Stratified bootstrap}
Let
\(
n_{\min}
=
\max(1,\lfloor n_{\text{boot}}/2\rfloor)
\)
and
\(
n_{\max}
=
\max(n_{\min},\lfloor\alpha n_{\text{pool}}\rfloor).
\)
Bootstrap sizes
\(n_{\text{boot}}^{(i)}\)
are allocated according to
\(\tilde{\sigma}_i^2\),
subject to
\(
n_{\min}\le n_{\text{boot}}^{(i)}\le n_{\max}
\)
and
\(
\sum_i n_{\text{boot}}^{(i)}
=
n\,n_{\text{boot}}.
\)
Each player first receives \(n_{\min}\) samples, after which the remaining budget is distributed proportionally, repeatedly redistributing residuals from capped players. Deterministic largest-remainder rounding produces integer allocations. Thus, higher-variance players receive more samples without increasing the total bootstrap budget. If \(\mathcal K=\emptyset\) or all players have identical scores, we instead set
\(
n_{\text{boot}}^{(i)}
=
n_{\text{boot}}
\)
for all players. In each game, player \(i\) bootstraps \(n_{\text{boot}}^{(i)}\) observations from \(\mathcal S_i\) to form \(\mathcal B_i^{(g)}\); coalition evaluation and Shapley estimation then proceed exactly as in the pooled method.

\end{revisionblock}

\begin{algorithm}[!t]
\caption{Stratified Pooled Method for Shapley Value Estimation}
\label{alg:stratified}
\begin{algorithmic}[1]
\REQUIRE Player distributions \(\{P_i\}_{i=1}^{n}\), pool size \(n_{\text{pool}}\), base bootstrap size \(n_{\text{boot}}\), allocation cap \(\alpha\), number of games \(n_{\text{games}}\), and permutations per game \(n_{\text{iter}}\)
\ENSURE Estimated Shapley values \(\{\phi_i\}_{i=1}^{n}\)
\STATE Initialize \(\phi_i\gets0\) for all \(i\)
\FOR{\(i\gets1\) to \(n\)}
    \STATE \(\mathcal{S}_i\gets\{(x_j,y_j)\sim P_i\}_{j=1}^{n_{\text{pool}}}\)
\ENDFOR
\STATE \(\mathcal{S}_{\text{all}}\gets\operatorname{concat}(\mathcal{S}_1,\ldots,\mathcal{S}_n)\)
\STATE \(\mathcal{K}\gets\{k:\operatorname{Var}(\mathcal{S}_{\text{all}}^{(k)})>0\}\)
\STATE \(\sigma_i^2\gets\sum_{k\in\mathcal{K}}\operatorname{Var}(\mathcal{S}_i^{(k)})/\operatorname{Var}(\mathcal{S}_{\text{all}}^{(k)})\) for all \(i\)
\STATE \(n_{\text{min}}\gets\max(1,\lfloor n_{\text{boot}}/2\rfloor)\), \(n_{\text{max}}\gets\max(n_{\text{min}},\lfloor\alpha n_{\text{pool}}\rfloor)\)
\IF{\(\mathcal{K}=\emptyset\) or \(\max_i\sigma_i^2=\min_i\sigma_i^2\)}
    \STATE \(n_{\text{boot}}^{(i)}\gets n_{\text{boot}}\) for all \(i\)
\ELSE
    \STATE \(\tilde{\sigma}_i^2\gets(\sigma_i^2-\min_j\sigma_j^2)/(\max_j\sigma_j^2-\min_j\sigma_j^2)\) for all \(i\)
    \STATE Allocate \(n_{\text{boot}}^{(i)}\) according to \(\tilde{\sigma}_i^2\), subject to \(n_{\text{min}}\leq n_{\text{boot}}^{(i)}\leq n_{\text{max}}\) and \(\sum_i n_{\text{boot}}^{(i)}=n n_{\text{boot}}\)
\ENDIF
\FOR{\(g\gets1\) to \(n_{\text{games}}\)}
    \FOR{\(i\gets1\) to \(n\)}
        \STATE \(\mathcal{B}_i^{(g)}\gets\operatorname{sample}(\mathcal{S}_i,n_{\text{boot}}^{(i)},\operatorname{replace}=\operatorname{True})\)
    \ENDFOR
    \STATE \(\{\bar{\phi}_i^{(g)}\}_{i=1}^{n}\gets\textsc{MC-Shapley}(\{\mathcal{B}_i^{(g)}\}_{i=1}^{n},n_{\text{iter}})\)
    \STATE \(\phi_i\gets\phi_i+\bar{\phi}_i^{(g)}\) for all \(i\)
\ENDFOR
\RETURN \(\{\phi_i/n_{\text{games}}\}_{i=1}^{n}\)
\end{algorithmic}
\end{algorithm}

\begin{table*}[!t]
\centering
\caption{Variance on Wine Quality under a \(2{,}500\)-record source-access cap, averaged across players and five replications. The baseline uses \(G_{\text{B}}=4\) fresh-data games; \(G_{\text{P/S}}\) denotes the pooled and stratified game count.}
\label{tab:fixed_source_variance_wine}
\resizebox{0.75\textwidth}{!}{%
\begin{tabular}{
S[table-format=4.0]
S[table-format=1.0]
S[table-format=3.0]
| S[table-format=1.2e-1] S[table-format=1.2e-1] S[table-format=1.2e-1]
| S[table-format=1.2e-1] S[table-format=1.2e-1] S[table-format=1.2e-1]
}
\toprule
{\(n_{\text{iter}}\)} &
{\(G_{\text{B}}\)} &
{\(G_{\text{P/S}}\)} &
\multicolumn{3}{|c|}{\(\overline{\mathrm{Var}(\widehat{\mathbb{E}}[\phi])}\)} &
\multicolumn{3}{c}{\(\overline{\mathrm{Var}(\widehat{\mathrm{Var}}(\phi))}\)} \\
& & &
\multicolumn{1}{|c}{Baseline} &
{Pooled} &
\multicolumn{1}{c|}{Stratified} &
{Baseline} &
{Pooled} &
{Stratified} \\
\midrule
100  & 4 & 50  & 3.51e-4 & 3.18e-5 & 3.24e-5 & 6.43e-7 & 5.00e-8 & 5.15e-8 \\
100  & 4 & 200 & 3.51e-4 & 1.46e-5 & 3.45e-5 & 6.43e-7 & 1.42e-8 & 1.93e-8 \\
1000 & 4 & 50  & 2.03e-4 & 2.02e-5 & 3.90e-5 & 2.88e-7 & 2.02e-8 & 1.94e-8 \\
1000 & 4 & 200 & 2.03e-4 & 2.09e-5 & 2.71e-5 & 2.88e-7 & 5.23e-9 & 7.22e-9 \\
\bottomrule
\end{tabular}%
}
\end{table*}

\begin{figure}[!t]
\centering
\includegraphics[width=\linewidth]{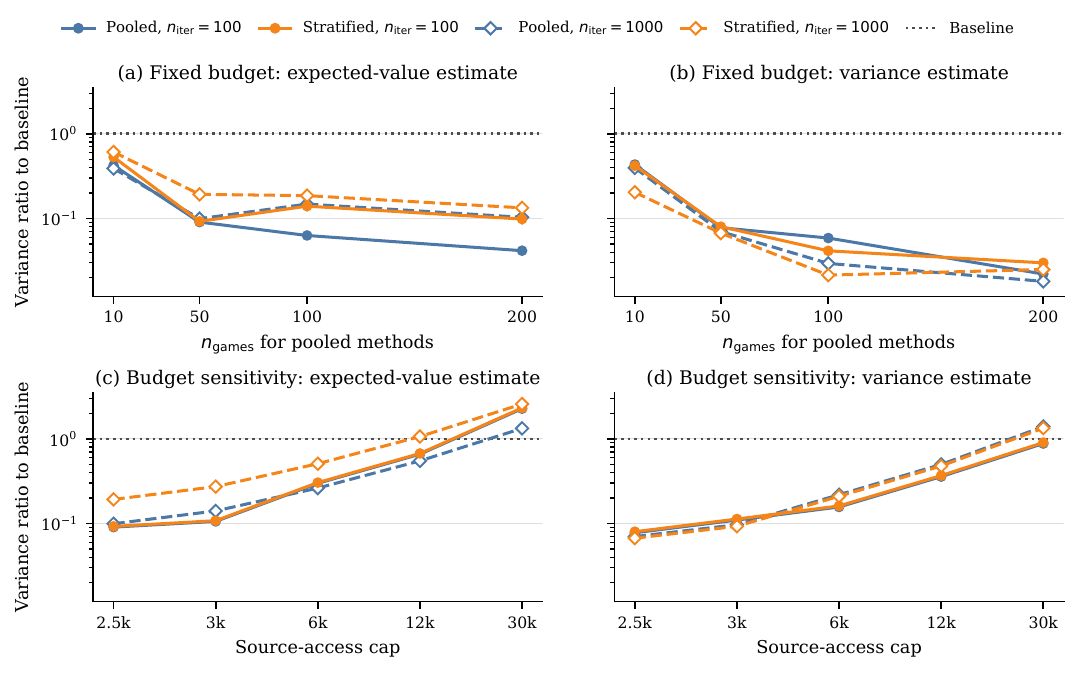}
\caption{Relative variance on Wine Quality. Top: effect of pooled-game count under a fixed source-access cap. Bottom: effect of the source-access budget.}
\label{fig:stratified_comparison}
\end{figure}

\begin{revisionblock}
\subsection{Variance Efficiency under Source-Sampling Constraints}
\label{sec:source-budget-variance}

We fix \(n_{\text{pool}}=250\), \(n_{\text{boot}}=60\), and \(\alpha=0.5\), and evaluate both variance metrics over five replications. Under a source-access budget of \(2{,}500\) records, the baseline completes four games using fresh samples, whereas the pooled methods access \(2{,}477\) records once and generate additional games through local resampling. We report the across-replication variances of the estimated per-player expectation and variance, averaged across players.

Under the fixed source-access budget, both pooled estimators substantially reduce both variance metrics by supporting additional games through local resampling. Table~\ref{tab:fixed_source_variance_wine} summarizes the results. At \(n_{\text{games}}=50\), they reduce \(\overline{\mathrm{Var}(\widehat{\mathbb{E}}[\phi])}\) by \(80.7\%\)--\(90.9\%\) and \(\overline{\mathrm{Var}(\widehat{\mathrm{Var}}(\phi))}\) by \(92.0\%\)--\(93.3\%\). At \(n_{\text{games}}=200\), the reductions increase to \(86.6\%\)--\(95.8\%\) and \(97.0\%\)--\(98.2\%\), respectively. Thus, pool reuse exchanges inexpensive local computation for substantially more stable Shapley estimates without additional source access.

Figure~\ref{fig:stratified_comparison} further shows that, under a fixed source-access budget, increasing the number of locally resampled games generally reduces both variance metrics, with all pooled configurations outperforming the baseline (top panels). As the source-access budget increases (bottom panels), the pooled methods remain superior through \(6{,}000\) records, and most gains persist at \(12{,}000\). At \(30{,}000\) records, the baseline supports enough fresh-data games to narrow or eliminate the advantage. Pooling is therefore most effective when source access is limited or expensive.
\end{revisionblock}

\begin{revisionblock}

\subsection{Robustness, Sensitivity, and Scalability}

\subsubsection{Generalization Across Datasets and Learning Models}

\paragraph{Dataset and utility function}
We further evaluate all three methods on the scikit-learn Digits dataset~\cite{pedregosa2011scikit}, derived from the UCI Optical Recognition of Handwritten Digits dataset~\cite{uci_optical_digits}, which contains \(1{,}797\) grayscale images from ten classes, each represented by \(64\) pixel-intensity features. We scale the features to \([0,1]\) and use a fixed stratified \(70/30\) train--test split for all methods, games, and replications.

We treat the ten digit classes as ten players, where player \(i\) contains the training samples labeled \(i-1\). Each nonempty coalition is formed by combining the corresponding classes. For every coalition, we train a classifier from scratch and define \(v(S)\) as its classification accuracy on the fixed test set, with \(v(\emptyset)=0\). Bootstrap duplicates are retained in the coalition training data.

The classifier consists of a \texttt{StandardScaler} followed by an MLP with hidden-layer sizes \(128\) and \(64\), ReLU activations, the L-BFGS optimizer~\cite{liu1989lbfgs}, regularization \(10^{-4}\), at most \(300\) iterations, and random seed \(314159\). The architecture, hyperparameters, split, and seed are fixed before Shapley estimation. Trained on the complete training set, the model achieves \(96.48\%\) test accuracy, confirming that the utility represents a meaningful multiclass task.

\paragraph{Variance reduction under a fixed source-sampling budget}

We compare the methods under a common source-access budget of \(3{,}000\) data draws. The baseline draws \(10\times60=600\) fresh samples per game and therefore evaluates five games, whereas the pooled and stratified methods each construct one pool of \(10\times300=3{,}000\) samples and reuse it across \(50\) locally resampled games.

The baseline achieves
\(\overline{\mathrm{Var}}(\widehat{\mathbb{E}}[\phi])=4.220\times10^{-7}\)
and
\(\overline{\mathrm{Var}}(\widehat{\mathrm{Var}}(\phi))=4.510\times10^{-12}\).
Under the same source budget, the pooled method reduces these metrics by \(53.1\%\) and \(74.5\%\), respectively, while the stratified method further improves the reductions to \(56.7\%\) and \(77.9\%\).

These results show that pooling improves estimation stability per source-data draw by reusing sampled data across additional games, with stratified allocation providing a further improvement. Because the comparison fixes source access rather than utility evaluations, it is most relevant when repeated access to the original data sources is the primary bottleneck.

\subsubsection{Hyperparameter Selection and Sensitivity}
\label{sec:hyperparameter_sensitivity}

\begin{figure}[!t]
\centering
\includegraphics[width=\linewidth]{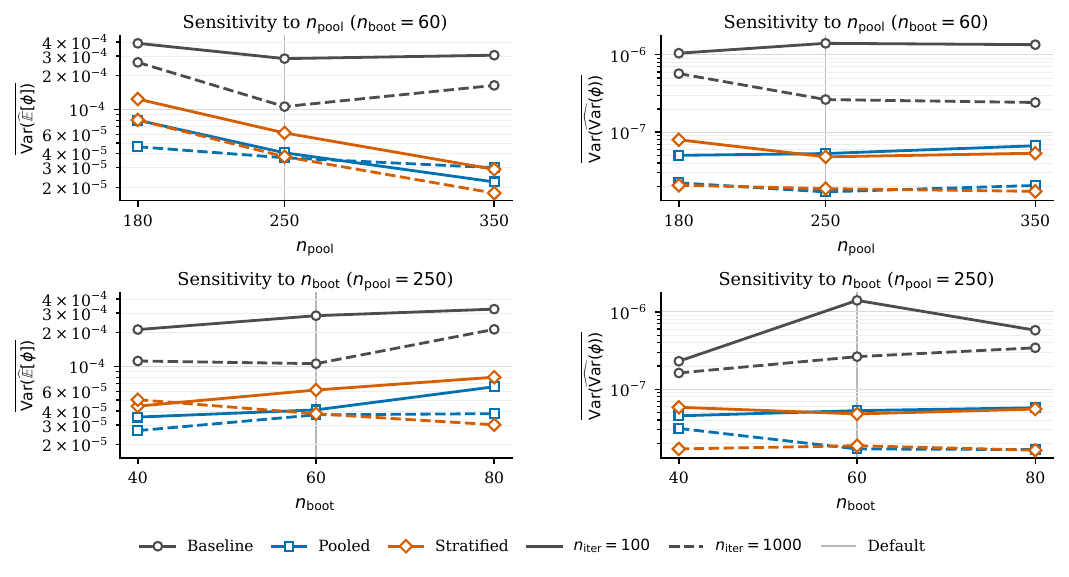}
\caption{Sensitivity to \(n_{\text{pool}}\) and \(n_{\text{boot}}\) on Wine Quality. The vertical axes report the two variance metrics on logarithmic scales. Solid and dashed lines denote \(n_{\text{iter}}=100\) and \(1000\), respectively; gray vertical lines mark the reference settings \(n_{\text{pool}}=250\) and \(n_{\text{boot}}=60\).}
\label{fig:wine_sensitivity}
\end{figure}

\paragraph{Selection rationale}
The three hyperparameters control different cost--quality trade-offs. The pool size \(n_{\text{pool}}\) determines one-time source access and pool representativeness, \(n_{\text{boot}}\) controls the number of records representing each player in a realized game, and \(\alpha\) limits the maximum player-specific allocation. We use \(n_{\text{pool}}=250\), \(n_{\text{boot}}=60\), and \(\alpha=0.5\) as the reference configuration. The pool is approximately \(4.2\) times the per-game bootstrap size, balancing source access and model-training cost. Moreover, \(n_{\text{pool}}=250\) and \(n_{\text{boot}}=60\) are interior points of the evaluated ranges rather than values selected to optimize either metric.

\paragraph{Sensitivity to pool and bootstrap sizes}
Figure~\ref{fig:wine_sensitivity} varies \(n_{\text{pool}}\in\{180,250,350\}\) with \(n_{\text{boot}}=60\) and \(n_{\text{boot}}\in\{40,60,80\}\) with \(n_{\text{pool}}=250\), using a source-access-matched baseline for every configuration. Across all settings and both \(n_{\text{iter}}\in\{100,1000\}\), the pooled and stratified methods reduce \(\overline{\mathrm{Var}}(\widehat{\mathbb{E}}[\phi])\) by \(54.9\%\)--\(92.6\%\) and \(\overline{\mathrm{Var}}(\widehat{\mathrm{Var}}(\phi))\) by \(74.6\%\)--\(96.6\%\) relative to the baseline. Increasing \(n_{\text{pool}}\) generally improves expected-value stability; however, increasing it from \(250\) to \(350\) increases realized source access by \(32.3\%\), from \(2{,}477\) to \(3{,}277\) records, without consistently improving the second metric. The effect of \(n_{\text{boot}}\) is also non-monotone, and no tested value uniformly minimizes both metrics. Thus, \(n_{\text{pool}}=250\) and \(n_{\text{boot}}=60\) provide reasonable intermediate settings rather than a universally optimal configuration. The observed reductions persist throughout the tested neighborhood, indicating that the conclusions are not sensitive to the reference settings. Stratified allocation remains competitive with uniform pooling and more frequently achieves the lower \(\overline{\mathrm{Var}}(\widehat{\mathrm{Var}}(\phi))\) at \(n_{\text{iter}}=1000\).

\paragraph{Sensitivity to the allocation cap}
We also examine the effect of \(\alpha\) under the reference configuration. With \(n_{\text{pool}}=250\), \(n_{\text{boot}}=60\), and \(\alpha=0.5\), the allocation cap is \(n_{\max}=\lfloor\alpha n_{\text{pool}}\rfloor=125\). Across all five replications and both permutation budgets, the largest provisional allocation is \(94\), so the cap is inactive and increasing \(\alpha\) can only relax it further. Consequently, every \(\alpha\in[0.5,1]\) yields identical allocations and results for these realized pools. We therefore choose the smallest value, \(\alpha=0.5\), as a conservative safeguard against extreme player-specific allocations rather than tuning it to optimize the reported metrics. The cap may become active when \(n_{\text{pool}}\) is small relative to \(n_{\text{boot}}\), so \(\alpha\) remains relevant outside the evaluated regime.

\subsubsection{Scalability Across Player Counts}
\label{sec:player-scalability}

We evaluate the baseline, pooled, and stratified methods at \(n\in\{10,20,50,100,200,500,1000\}\) using a heterogeneous additive game with exact references. Each player has a fixed empirical source of \(5{,}000\) Gaussian observations, with means linearly spaced over \([-3,3]\) and standard deviations geometrically spaced over \([0.1,2.0]\), randomly permuted to decouple location and dispersion. For \(u(C)=\sum_{i\in C}\bar{x}_i\), player \(i\)'s Shapley value equals its sampled mean. Thus, for the common 128-record target, the exact expectation is the empirical source mean and the exact variance is the empirical population variance divided by \(128\). The pooled variance estimate is finite-pool corrected, and the stratified estimate is normalized to the same 128-record target before comparison.

All methods observe \(512n\) source records. The baseline evaluates four games with 128 fresh records per player, whereas the pooled method accesses a 512-record pool per player once and constructs \(50\) local bootstrap games of size \(128\). The stratified method uses the same pools and games but allocates \(128n\) records per game according to scaled player-variance scores. With \(\alpha=0.5\), allocations range from \(64\) to \(256\) records per player while summing to \(128n\). Each game uses four permutations, and all results are averaged over ten independent replications.

Table~\ref{tab:player-scalability} reports the stability metrics \(\overline{\mathrm{Var}(\widehat{\mathbb{E}}[\phi])}\) and \(\overline{\mathrm{Var}(\widehat{\mathrm{Var}}(\phi))}\), computed as the sample variance across replications for each player and then averaged across players. Within replication \(r\), accuracy is measured by
\begin{equation}
\mathrm{MAE}_{\mathbb{E}}^{(r)}
=\frac{1}{n}\sum_{i=1}^{n}
\left|\widehat{\mathbb{E}}_{r}[\phi_i]-\mathbb{E}^{*}[\phi_i]\right|.
\label{eq:mae-expectation}
\end{equation}

\begin{equation}
\mathrm{MAE}_{\mathrm{Var}}^{(r)}
=\frac{1}{n}\sum_{i=1}^{n}
\left|\widehat{\mathrm{Var}}_{r}(\phi_i)-\mathrm{Var}^{*}(\phi_i)\right|.
\label{eq:mae-variance}
\end{equation}

The reported MAEs average these quantities over the ten replications.

Table~\ref{tab:player-scalability} presents representative player counts, while the following ranges summarize all seven values of \(n\). None of the four metrics deteriorates systematically as \(n\) increases to \(1{,}000\). Across the full sweep, the expectation MAEs of the pooled and stratified methods differ from the baseline by at most \(4.2\%\), with paired \(95\%\) confidence intervals including zero at every \(n\). Their \(\overline{\mathrm{Var}(\widehat{\mathbb{E}}[\phi])}\) values remain between \(0.85\) and \(1.13\) times those of the baseline.

The improvements are larger for Shapley-variance estimation. The pooled method reduces \(\mathrm{MAE}_{\mathrm{Var}}\) by \(69.0\%\)--\(75.4\%\) and \(\overline{\mathrm{Var}(\widehat{\mathrm{Var}}(\phi))}\) by \(88.5\%\)--\(94.4\%\), while the stratified method achieves reductions of \(64.6\%\)--\(77.9\%\) and \(81.0\%\)--\(96.8\%\), respectively. The confidence intervals for all variance-MAE improvements exclude zero, although neither pool-based method uniformly dominates the other.

All measured costs scale linearly with \(n\). The baseline uses \(4n\) source requests and \(16n\) utility evaluations, whereas each pool-based method uses \(n\) source requests and \(200n\) utility evaluations while observing the same \(512n\) source records. Thus, pooling reduces source requests by \(75\%\) while preserving expectation accuracy and improving variance accuracy and stability, at the cost of additional local computation.
\end{revisionblock}

\begin{table}[t]
\centering
\caption{Effectiveness under an equal source-record budget at representative player counts. The two variance metrics measure across-replication stability averaged over players; MAEs are computed against the exact 128-record references and averaged over players and ten replications. Lower is better.}
\label{tab:player-scalability}
\resizebox{\columnwidth}{!}{%
\begin{tabular}{@{}rl*{4}{S[table-format=1.2e-1]}@{}}
\toprule
{\(n\)} & {Method}
& {\(\overline{\mathrm{Var}(\widehat{\mathbb{E}}[\phi])}\)}
& {\(\overline{\mathrm{Var}(\widehat{\mathrm{Var}}(\phi))}\)}
& {\(\mathrm{MAE}_{\mathbb{E}}\)}
& {\(\mathrm{MAE}_{\mathrm{Var}}\)} \\
\midrule
10
& Baseline   & 2.08e-3 & 6.51e-5 & 2.73e-2 & 4.13e-3 \\
& Pooled     & 2.11e-3 & 7.47e-6 & 2.63e-2 & 1.28e-3 \\
& Stratified & 2.35e-3 & 1.24e-5 & 2.75e-2 & 1.46e-3 \\
\addlinespace[1.5pt]
100
& Baseline   & 1.46e-3 & 6.49e-5 & 2.37e-2 & 3.41e-3 \\
& Pooled     & 1.29e-3 & 3.66e-6 & 2.27e-2 & 8.39e-4 \\
& Stratified & 1.27e-3 & 4.47e-6 & 2.28e-2 & 9.28e-4 \\
\addlinespace[1.5pt]
1000
& Baseline   & 1.31e-3 & 5.58e-5 & 2.24e-2 & 3.25e-3 \\
& Pooled     & 1.39e-3 & 3.87e-6 & 2.31e-2 & 8.84e-4 \\
& Stratified & 1.35e-3 & 3.74e-6 & 2.29e-2 & 8.75e-4 \\
\bottomrule
\end{tabular}%
}
\end{table}

\begin{revisionblock}
\subsection{Measured Runtime under Simulated Source Latency}
\label{sec:measured-efficiency}

Reducing fresh-source access does not necessarily reduce wall time proportionally because source requests can be batched and local utility evaluation may dominate runtime. We therefore measure source-access and end-to-end runtime under simulated per-request latency. A \emph{batched source request} retrieves all samples required from one player in a single call. The baseline makes \(n\,n_{\text{games}}\) requests, whereas each pooled estimator makes only \(n\). The emulator injects a delay of 0, 10, or 50 ms per request, independent of batch size; measured access time also includes sampling and copying.

Let \(T_{\text{access}}\) denote cumulative source-access time, \(T_{\text{wall}}\) the end-to-end runtime, and \(T_{\text{nonaccess}}=T_{\text{wall}}-T_{\text{access}}\) the remaining local computation. For Wine Quality, we use \(n=10\), \(n_{\text{games}}=5\), \(n_{\text{iter}}=8\), \(n_{\text{sample}}=n_{\text{boot}}=32\), \(n_{\text{pool}}=128\), and \(\alpha=0.75\). This reduced configuration isolates the effect of source latency rather than representing the full experimental setting. All experiments use five matched seeds and single-threaded execution on the same Apple M1 Pro CPU. Table~\ref{tab:measured-efficiency} reports the mean and sample standard deviation over five runs; speedup is computed for each paired seed before averaging.

\begin{figure}[!t]
\centering
\includegraphics[width=\linewidth]{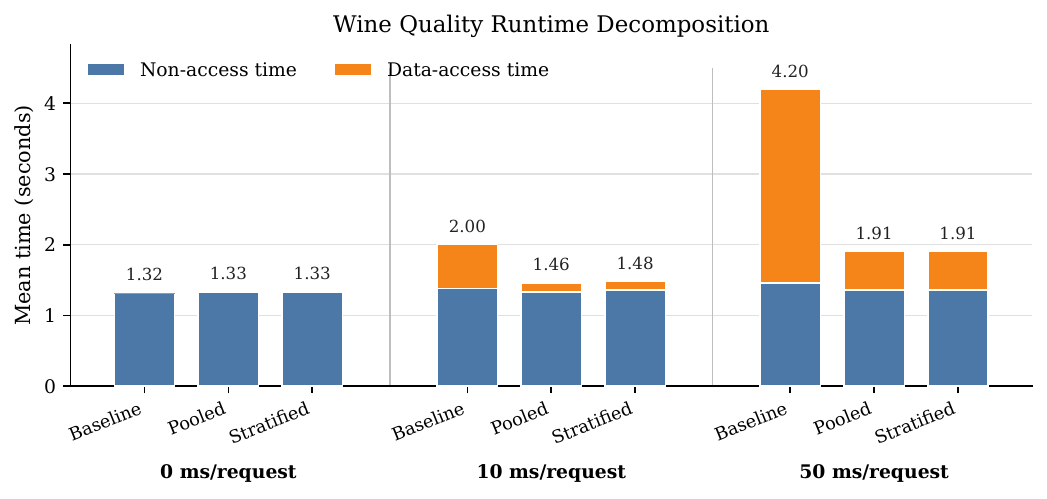}
\caption{Mean runtime decomposition across five seeds. Stacked bars show source-access and non-access time, with total height equal to end-to-end wall time.}
\label{fig:runtime_efficiency}
\end{figure}

\begin{table}[!t]
\centering
\caption{Runtime under simulated per-request source latency. Times are mean \(\pm\) sample standard deviation over five seeds, in seconds. Speedup is the mean wall-time ratio relative to baseline.}
\label{tab:measured-efficiency}
\scriptsize
\setlength{\tabcolsep}{2.5pt}
\resizebox{\columnwidth}{!}{%
\begin{tabular}{r l r c c c c}
\toprule
Latency & Method & Requests & \(T_{\text{access}}\) & \(T_{\text{nonaccess}}\) & \(T_{\text{wall}}\) & Speedup \\
(ms/req.) & & & (s) & (s) & (s) & \\
\midrule
0 & Baseline   & 50 & \(0.00149\pm0.00004\) & \(1.323\pm0.011\) & \(1.324\pm0.011\) & \(1.000\times\) \\
0 & Pooled     & 10 & \(0.00033\pm0.00003\) & \(1.330\pm0.012\) & \(1.331\pm0.012\) & \(0.995\times\) \\
0 & Stratified & 10 & \(0.00031\pm0.00002\) & \(1.330\pm0.009\) & \(1.330\pm0.009\) & \(0.995\times\) \\
\midrule
10 & Baseline   & 50 & \(0.6222\pm0.0043\) & \(1.379\pm0.017\) & \(2.002\pm0.016\) & \(1.000\times\) \\
10 & Pooled     & 10 & \(0.1245\pm0.0027\) & \(1.333\pm0.027\) & \(1.458\pm0.028\) & \(1.374\times\) \\
10 & Stratified & 10 & \(0.1222\pm0.0015\) & \(1.358\pm0.044\) & \(1.480\pm0.045\) & \(1.353\times\) \\
\midrule
50 & Baseline   & 50 & \(2.7345\pm0.0102\) & \(1.462\pm0.019\) & \(4.196\pm0.027\) & \(1.000\times\) \\
50 & Pooled     & 10 & \(0.5446\pm0.0074\) & \(1.361\pm0.020\) & \(1.906\pm0.016\) & \(2.202\times\) \\
50 & Stratified & 10 & \(0.5453\pm0.0073\) & \(1.361\pm0.008\) & \(1.906\pm0.013\) & \(2.202\times\) \\
\bottomrule
\end{tabular}%
}
\end{table}

Pooling reduces source requests from 50 to 10 (80\%) in the controlled benchmark. Figure~\ref{fig:runtime_efficiency} shows that non-access computation is nearly unchanged across methods, whereas the wall-time benefit grows with source latency. At 0 ms/request, all methods require 1.324--1.331 s, so pooling provides essentially no speedup. At 10 ms/request, pooling reduces access time from 0.6222 to 0.1245 s and wall time from 2.002 to 1.458 s, corresponding to a 27.2\% reduction and a \(1.374\times\) speedup. At 50 ms/request, access time decreases from 2.7345 to 0.5446 s and wall time from 4.196 to 1.906 s, yielding a 54.6\% reduction and a \(2.202\times\) speedup. Stratified pooling achieves comparable gains, indicating that the improvement is primarily due to reduced source access.

\subsection{Scope of Empirical Comparison}
\label{sec:comparison-scope}

Existing uncertainty-aware Shapley frameworks do not estimate our target. Distributional Shapley assigns expected marginal values to individual data points relative to a background distribution~\cite{ghorbani2020distributional,kwon2021aistats}; U-statistic inference quantifies uncertainty in individual-data Shapley estimation~\cite{wu2024uncertainty}; and uncertain-value and Gaussian-process Shapley methods propagate uncertainty in utilities or predictive models to feature attributions~\cite{heese2023shapley,chau2023explaining}. None models providers as players, repeatedly samples fixed-size datasets from provider distributions, and estimates both the expectation and variance of the resulting provider-level Shapley value. Applying these methods would therefore change the players, the randomness, or the estimand. We therefore use the fresh-sampling baseline as the common-target reference and compare the pooled estimators under matched source-access budgets. This defines the scope of our empirical comparison rather than claiming superiority over these complementary frameworks.

\end{revisionblock}

\section{Related Work}\label{sec:related}

The Shapley value, introduced by Shapley~\cite{shapley1953value}, provides an axiomatic framework for fair value allocation in cooperative games. It has since been applied to economics, political science~\cite{shapley1954method,winter2002shapley}, and, more recently, machine learning for data valuation~\cite{ghorbani2019data,pei2020survey,lin2024comprehensive}. We review related work on data-Shapley models and Shapley-value computation.

\subsection{Models of Data Shapley}

Ghorbani and Zou~\cite{ghorbani2019data} introduced \emph{data Shapley}, which measures the contribution of each training sample to model performance. Their formulation assumes a fixed dataset and has inspired extensive follow-up work~\cite{jia2019towards,pei2020survey}.

\revise{Ghorbani \textit{et al.}~\cite{ghorbani2020distributional} subsequently proposed \emph{distributional Shapley}, which values a data point relative to an underlying data distribution, and Kwon \textit{et al.}~\cite{kwon2021aistats} developed efficient algorithms for this point-level estimand. These methods contextualize the value of individual data points but do not estimate the variance of provider-level Shapley values across repeated fixed-size samples from provider-specific distributions.}

Recent work has explored alternative valuation models. Liu \textit{et al.}~\cite{liu20232d} proposed 2D-Shapley for fragmented data sources, Wang \textit{et al.}~\cite{wang2024data} introduced in-run data Shapley for single training runs, and Garrido Lucero \textit{et al.}~\cite{garrido2024shapley} developed DU-Shapley as an efficient proxy estimator. These methods address different valuation settings or computational objectives rather than repeated provider-level sampling.

\revise{Wu \textit{et al.}~\cite{wu2024uncertainty} related individual-data Shapley estimation to infinite-order U-statistics and derived confidence intervals. Shapley values with uncertain value functions~\cite{heese2023shapley} model uncertainty in the utility, whereas stochastic Shapley values for Gaussian processes propagate predictive uncertainty to feature attributions~\cite{chau2023explaining}. Although these methods quantify meaningful forms of uncertainty, their players and estimands differ from provider-level Shapley values induced by repeated sampling from provider distributions. Section~\ref{sec:comparison-scope} therefore defines the scope of our empirical comparison.}

\subsection{Data Shapley Computation}

\revise{Computing the exact Shapley value is \#P-hard in general because it requires evaluating all coalitions~\cite{deng1994complexity}. Consequently, extensive work has studied sampling-based approximation. Castro \textit{et al.}~\cite{castro2009random} proposed polynomial-time permutation sampling for fixed cooperative games. Maleki \textit{et al.}~\cite{maleki2013bounding} established non-asymptotic error bounds and studied stratified sampling, while Castro \textit{et al.}~\cite{castro2017improving} further improved the estimator through optimum-allocation stratified sampling. Subsequent work has explored ergodic sampling~\cite{illes2019ergodic}, adaptive stratified sampling~\cite{burgess2021bernstein,zhang2023efficient}, variance reduction~\cite{wu2023variance}, and structured permutation sampling~\cite{mitchell2022sampling}.}

\revise{For data valuation, Ghorbani and Zou~\cite{ghorbani2019data} proposed \emph{Truncated Monte Carlo Shapley} and \emph{Gradient Shapley}; Jia \textit{et al.}~\cite{jia2019towards} developed KNN-based estimators; and Pang \textit{et al.}~\cite{pang2025shapley} reduced estimation variance through differential-matrix reconstruction.}

Other methods exploit structural properties of the utility function. Luo \textit{et al.}~\cite{luo2024fast} studied binary-utility data assemblage tasks using cooperative simple games. Zhang \textit{et al.}~\cite{zhang2023dynamic} and Xia \textit{et al.}~\cite{xia2025computing} developed incremental algorithms for dynamic datasets. Luo and Pei~\cite{luo2024applications} provide a comprehensive survey of Shapley-value applications and computation.

\revise{Existing methods substantially improve the computational efficiency of Shapley estimation but generally assume a fixed cooperative game or contributed dataset. In particular, existing stratified methods allocate computational samples across permutations, coalition sizes, or marginal contributions to estimate a fixed Shapley value. In contrast, our framework models each participant by a probabilistic data distribution, estimates both the expectation and variance of the resulting Shapley value across data realizations, and reallocates the data-sampling budget across players according to their observed variability rather than across permutations or coalitions.}

\section{Conclusion}
\label{sec:conclusion}

\begin{revisionblock}
We studied Shapley value estimation when participants contribute random samples from underlying data distributions. We characterized the expectation and variance of the resulting Shapley values, derived unbiased outer-sample estimators when realized-game values are computed exactly, and developed a fresh-sampling baseline together with pooled and stratified pooled estimators. Experiments on synthetic and real datasets showed that the pooled methods substantially improve estimation stability under constrained source access while preserving expectation accuracy, highlighting an accuracy--source-access trade-off rather than uniform superiority over fresh sampling. Future work will extend this framework to federated learning with randomized client participation, where participation uncertainty introduces an additional source of randomness in data valuation.
\end{revisionblock}

\bibliographystyle{IEEEtran}
\bibliography{references}



\begin{IEEEbiographynophoto}{Zhuofan Jia}
is a Ph.D. candidate at Duke University. Her research interests include machine learning, data science, and artificial intelligence.
\end{IEEEbiographynophoto}

\begin{IEEEbiographynophoto}{Jian Pei} is the Arthur S. Pearse Distinguished Professor at Duke University. His research interests include data mining, machine learning, artificial intelligence, and data science, with a focus on developing scalable methods for knowledge discovery and data analytics.
\end{IEEEbiographynophoto}

\end{document}